# Recent advances in label-free imaging and quantification techniques for the study of lipid droplets in cells

Hyeonwoo Kim[1], Seungeun Oh[2], Seongsoo Lee[3,4], Kwang suk Lee[5] and YongKeun Park[6,7,8,*]

[1]Department of Biological Sciences, Korea Advanced Institute of Science and Technology (KAIST), Daejeon 34141, Republic of Korea

[2]Department of Physics and Cellular Molecular Medicine, University of California, San Diego, California 2093, USA

[3]Gwangju Center, Korea Basic Science Institute (KBSI), Gwangju 61751, Republic of Korea

[4]Department of Systems Biotechnology, Chung-Ang University Anseong-si, Gyeonggi-do 17546, Republic of Korea

[5]Department of Urology, Yonsei University College of Medicine, Gangnam Severance Hospital, Seoul 06229, Republic of Korea

[6]Department of Physics, KAIST, Daejeon 34141, Republic of Korea

[7]KAIST Institute for Health Science and Technology, KAIST, Daejeon 34141, Republic of Korea

[8]Tomocube Inc., Daejeon 34109, Republic of Korea

[*]E-mails: yk.park@kaist.ac.kr

**Abstract**

Lipid droplets (LDs), once considered mere storage depots for lipids, have gained recognition for their intricate roles in cellular processes, including metabolism, membrane trafficking, and disease states like obesity and cancer. This review explores label-free imaging techniques' applications in LD research. We discuss holotomography and vibrational spectroscopic microscopy, emphasizing their potential for studying LDs without molecular labels, and we highlight the growing integration of artificial intelligence. Clinical applications in disease diagnosis and therapy are also considered.

## 1. Introduction

As subcellular organelles, Lipid droplets (LDs) consist of a hydrophobic core of neutral lipids such as triacylglycerols (TGs) and sterol esters surrounded by a monolayer of phospholipids. LDs are crucial for maintaining lipid homeostasis in cells, and its formation is regulated by cellular levels of lipids. Cellular accumulation of lipids drives synthesis of triacylglycerols and sterol esters in ER, which eventually forms lens of oil [1]. By that formation, free fatty acids get stored in the lens and then cells avoid lipotoxicity in high lipid condition. The growth of the oil lens leads to budding of the droplet into the cytosols as LD. At low concentration of lipids or at the energy demanding condition, neutral lipids are hydrolyzed and LDs get smaller and eventually disappear (1).

Traditional methods for imaging lipid droplets, such as fluorescence microscopy, often require the use of molecular labels, which can interfere with cellular function and limit dynamic studies. Furthermore, these conventional techniques often lack the capability to deliver quantitative data. Recently, label-free imaging techniques have garnered attention for their ability to circumvent these limitations. Methods such as holotomography (HT), Raman microscopy, and Coherent Anti-Stokes Raman Scattering (CARS) microscopy have emerged as powerful tools for the high-resolution imaging and quantification of lipid droplets in living cells. This review aims to explore these cutting-edge label-free techniques, emphasizing their technological underpinnings, capabilities, and limitations.

In a rapidly evolving field, this article serves as a timely review of the current landscape, offering insights into how these technologies can further our understanding of lipid droplet biology. The subsequent sections will delve into each technique in detail, critically evaluating their contribution to advancing the field.

## 2. Current understanding lipid droplets in cells

LDs in adipocytes and hepatocytes are well characterized as these cells have high amount of lipids. However, other organs and even cancer cells also utilize LDs and recent research provides more knowledge with several roles (Fig. 1). Skeletal muscle cells have LDs for energy source, but LDs are also crucial for fate determination of skeletal muscle satellite cells; satellite cells having high LDs tend to differentiate while the cells having low LDs tend to proliferate [2]. Lipid metabolism is modified in several cancers, and LDs act for cancer cell survival and proliferation. Clear cell renal cell carcinoma is characterized by highly accumulated LDs and epigenetic regulation of diacylglycerol O-acyltransferase (DGAT1) is reported as a crucial for LDs formation [3]. In breast cancer, LD accumulation is cellular adaption to extracellular acidity in the malignant tumors and associated with poor overall survival [4].

The size of LDs has been considered to affect cellular functions, and the mechanism to explain how the size is regulated has been reported. Thermogenic adipocytes have multilocular LDs, which is more efficient to hydrolyze TGs than bigger LDs. Calsyntenin-3β (CLSTN3β) contributes to inhibits lipid transfer between LDs, thereby LD fusion and expansion are prevented in adipocytes [5] while IL-6 receptor signaling contributes to expansion of LDs in hepatocytes [6].

As subcellular organelles, LDs interact with mitochondria for supplement of fatty acids, but without clear mechanism. Recent study showed that fatty acid transporter 4 (FATP4) and Ras-related protein Rab-8A (Rab8a) on mitochondrial membrane is used to interact with Perilipin-5 (PLIN5) in LDs [7,8]. LDs are also interacting with lysosome for removal of TGs in macrophages and ADP-ribosylation factor-like 8b (Arl8b) in LDs mediates LD-lysosome contact [9]. Interestingly, another study showed that BCL2/adenovirus E1B 19 kDa protein-interacting protein 3 (BNIP3) contributes to colocalization of LDs, mitochondria, and lysosome for LD turnover [10].

## 3. Conventional methods to stain or label lipid droplets

An approach for the analysis of LD involves the application of transmission electron microscopy (TEM) which could be identified to selectively combine electron-rich osmium-based chemicals with unsaturated fatty acids [11]. However, in either chemical treated or fixed cell, it has distinct a limitation to provide insights into the physiological condition. Confocal fluorescence [12,13] or super-resolution microscopy [14,15] as supported an exceptional opportunity for the analysis of LD regulation employing LD-specific probes like as Nile Red, BODIPY and LipidTOX (Fig. 2). Furthermore, immunohistochemistry using confocal fluorescence microscopy serves as the primary approach for visualization of intracellular lipid droplets and the indirect detection of lipid droplet-associated proteins like seipin, PEX30 and perilipin (PLIN) family [16,17]. Nevertheless, it is important to note that this technique may impact intracellular activities due to potential issues like phototoxicity and photobleaching. In recent years, there has been a growing utilization of diverse instruments, including holotomography (HT) [18], coherent anti-Stokes Raman spectroscopy (CARS) [19], and Stimulated Raman Scattering (SRS) [20] for the purpose of imaging and monitoring living cells without the requirement of exogenous labelling agents. These methods in cellular imaging have played a pivotal role in advancing the field by providing non-invasive and label-free approaches for the comprehensive study of living cell structures and dynamics.

## 4. Holotomography

Holotomography (HT), also called optical diffraction tomography, is a label-free imaging method measuring the 3D refractive index (RI) distribution in live cells [21]. By combining tomographic reconstruction with quantitative phase imaging, HT reconstructs 3D RI maps of unlabeled cells from multiple 2D optical images with varying illumination. Unlike fluorescence, HT avoids chemical labels, eliminating phototoxicity and photobleaching. HT is widely used in cell biology, and regenerative medicine [22-25], due to its label-free, high-resolution, long-term, 3D, and quantitative imaging capabilities. Figure 3a shows the HT images of lipid droplets in live unlabeled Hep3B cells measured using a HT system (HT-X1, Tomocube Inc., South Korea).

The use of HT in researching lipid droplets provides unique advantages. HT enables real-time tracking of lipid droplet dynamics, including formation, growth, and interactions with other cellular components under physiologically relevant conditions. Since RI values of lipid droplets are significantly higher than those of other cellular organelles [26], straightforward thresholding allows for the segmentation of unlabeled cell lipid droplets. Also, a RI is linearly proportional to the biomolecular concentration [27]. Thus, HT's quantitative capabilities allow precise measurements of lipid droplet size, volume, concentration, and dry mass within cells, crucial for comprehending their physiological and pathological roles.

HT was used to measure and quantify the lipid droplets in unlabeled live hepatocytes, preadipocytes, and microalgae cells [26,28,29]. Time-lapse HT imaging was utilized to evaluated the therapeutic effects of a nanodrug designed to affect the targeted delivery of lobeglitazone to foam cell, by systematically quantifying the decreases of LDs in foam cells [18]. In a recent work, HT was utilized to investigate the roles of alpha-lipoic acid (α-LA) as a potential intervention for preventing hearing loss caused by cisplatin treatment [30].

## 5. Vibrational spectroscopic imaging

Vibrational spectroscopic imaging allows label-free chemical analysis for lipid droplets [31]. Lipids exhibit strong vibrational spectra in the 2800–3000 cm$^{-1}$ range which provide rich chemical information including abundance of TGs and cholesterol esters and the degree of saturation [32]. Confocal Raman microscopy can provides spatially resolved spectra of lipid droplets at a slow speed. Nonlinear Raman imaging, employing Coherent Anti-Stokes Raman Scattering (CARS) or Stimulated Raman Scattering (SRS) offers high-speed, high-sensitivity vibrational imaging in live biological samples with 3D resolution. CARS microscopy, which became commercially available, has been applied to studying lipid droplet dynamics and composition [33]. SRS, with lower background and superior quantification due to molecular orientation insensitivity, has been applied to analyzes dynamics and composition of lipid droplets [34]. Figure 3B presents the representative SRS images of lipid droplets in cells and tissues.

Raman imaging can combine with probes like fluorescent or Raman probes, utilizing unique Raman bands in silent regions. Minimally perturbing deuterium tag was employed in confocal Raman and SRS imaging to study lipid droplets [35], de novo lipogenesis [36], and lipids' role in ferroptosis [37].

HT intersects with chemical imaging through mid-infrared photothermal imaging (MIP). In MIP, mid-infrared light induces chemical bond-specific local heating, altering the refractive index, which can be detectable via quantitative phase imaging [38]. MIP can visualize lipid droplets using the carbon-hydrogen vibrational band (2800–3000 cm$^{-1}$) with lower phototoxicity than nonlinear Raman imaging.

## 6. Future applications and open questions

### 1. AI-enables segmentations

The advent of artificial intelligence (AI) has ushered in new possibilities for the study of lipid droplets through label-free imaging techniques. AI's computational prowess can augment the analytical capabilities of these techniques, leading to faster, more accurate, and more in-depth investigations [39].

Deep learning algorithms have shown promise in the segmentation of lipid droplets from label-free image data, enabling the precise identification and quantification of these organelles. These algorithms can be trained to differentiate lipid droplets from other cellular structures, providing detailed morphological data that can enhance our understanding of lipid droplet biology.

Moreover, machine learning models are increasingly being used to make predictive inferences about the interactions between lipid droplets and other subcellular organelles [40]. These models can analyze label-free imaging data to generate hypotheses on lipid droplet function, rate of formation, and degradation, among other aspects.

Real-time monitoring of lipid droplets becomes more feasible with AI, as automated data processing and analysis can occur simultaneously with data acquisition. This is particularly valuable for dynamic studies where changes in lipid droplet size, number, or localization might be indicative of cellular responses to various stimuli or conditions.

## 2. Unanswered questions in cell biology

Unsaturated fatty acids and saturated fatty acids are quite different in terms of lipid homeostasis. Unsaturated fatty acids induce TG synthesis and inhibit lipogenesis to lower cellular fatty acid levels while saturated fatty acids do not contribute to homeostasis [41], but raise lipotoxicity, which is involved in metabolic disorders, cardiovascular diseases, and inflammation [10]. The mechanism to explain the different outcomes is not yet revealed, and it is not clear whether the ratio of unsaturated and saturated fatty acids affect dynamics of LDs such as contact of LDs and mitochondria. LDs in cells are often shown with different size and dynamics, which indicates their own heterogeneity. Therefore, it is also crucial to understand how heterogenous LDs are in the same cells depending on location in cells, proximity to subcellular organelles, and intrinsic and extrinsic stimuli.

Previous research reveals that interactions between LDs and other organelles can shift from dynamic to stable, based on changing metabolic energy demands [42,43]. This highlights the potential of AI-Holotomography (HT), which unravels the complex dynamics of LD interactions with organelles. The integration of AI-HT with recognition algorithms and statistical techniques enhances the cost-effectiveness and accessibility of visualizing dynamic LDs in health and diseases. Furthermore, time-lapse monitoring and quantitative analysis of the dynamic LDs in live cells can be performed to evaluate whether the real-time quantitative analysis of LDs interacting other organelles is promising tool for developing new drugs against metabolic diseases.

## 3. Correlative with other techniques

Non-destructive nature of label-free imaging modalities is conducive to multiplexing with other methodologies. Correlative microscopy with electron microscopy, single-molecule imaging, spatial transcriptomics, and imaging mass spectrometry add rich molecular information to the quantitative and dynamic information acquired by label-free lipid droplet imaging, and label-free optical microscopy can provide complementary high-resolution information to omics data. Raman imaging with its rich chemical information has been fused with MALDI-mass spectrometry [44]. Simultaneous profiling of Raman spectra and single-cell transcriptome demonstrated that it is possible to infer transcriptomic profile from Raman spectra alone [45]. It may be possible that similar predictive power can be established for other label-free imaging modalities through correlative imaging [46].

## 4. Clinical applications

Cancers are marked by neovascularization and rapid cell proliferation, increasing energy demand. Cancer cells favor energy production even under aerobic conditions through reprogrammed metabolism. Dysregulated lipid metabolism leads to increased lipid content in vesicles, impacting cancer aggressiveness [47].

Investigations for the several energy metabolisms may lead to advancements in therapeutic interventions. In clinical points of view, obesity is a known factor in the development of several cancers, such as breast, colorectal, endometrial, kidney, esophageal, pancreatic, and liver cancer. Studies have shown the prognostic effects of obesity, energy metabolism, and lipid droplets in cancer. Recently, the therapeutic effect of combined treatments with chemotherapy and the drugs for energy metabolisms [Metformin as a first-line treatment for type 2 diabetes mellitus (T2DM) and Statins as 3-hydroxy-3-methylglutaryl coenzyme-A reductase (HMG-CoA reductase) inhibitor], have been investigated in various cancers [48,49].

Lipid droplet formation regulates cellular functions in various organs. Nonalcoholic fatty liver disease (NAFLD), a common chronic liver disease, results from metabolic stress and insulin resistance, causing hepatic steatosis due to excessive lipid accumulation in hepatocytes, leading to increased mitochondrial reactive oxygen species and fatty acid oxidation [50]. In chronic kidney disease (CKD), rising evidence links pathogenic lipid droplet buildup in kidney parenchyma to disease progression, highlighting its global health significance. [51] Further research is needed to understand how lipid buildup promotes kidney fibrosis and identify disease-specific triggers of lipid dysmetabolism.

In clinical practice, non-invasive methods like blood tests and imaging (ultrasound, CT, MRI) are recommended for disease diagnosis, prognosis, and monitoring. Biopsies, while invasive, can enhance diagnostic accuracy when combined with label-free optical monitoring, especially for diseases involving lipid droplets.

## 7. Conclusion

This review highlights recent advancements in label-free imaging and quantification techniques for studying LDs in cells. LDs, once considered simple lipid storage organelles, are now recognized for their vital roles in cellular processes and their implications in diseases like obesity and cancer. The review discusses the applications of label-free techniques such as HT, Raman microscopy, and CARS microscopy, emphasizing their potential to study LDs without the need for molecular labels. Additionally, each label-free technique can be coupled with other imaging modalities, providing multi-parametric insights into cellular function and lipid metabolism [52]. Given its non-invasive and quantitative capabilities, these label-free methods are poised to significantly advance our understanding of lipid biology in health and disease.

Furthermore, the integration of AI in image analysis offers promising avenues for studying LDs more comprehensively and efficiently. AI-driven segmentation and analysis can provide precise insights into LD morphology and function. These techniques have the potential to enhance our understanding of LD biology in various cellular contexts and diseases.

Additionally, the clinical applications of LD research are highlighted, especially in the context of cancer, obesity, NAFLD, and CKD. Label-free optical monitoring, combined with traditional clinical methods, offers non-invasive and potentially more informative approaches for disease diagnosis and prognosis. Overall, this review provides a comprehensive overview of the current state of label-free imaging techniques in LD research, emphasizing their significance in advancing our understanding of LD biology and their potential clinical applications.

**Acknowledgement**

This work was supported by National Research Foundation of Korea (2015R1A3A2066550, 2022M3H4A1A02074314), Institute of Information & communications Technology Planning & Evaluation (IITP; 2021-0-00745) grant funded by the Korea government (MSIT), the Korea Health Technology R&D Project through the Korea Health Industry Development Institute (KHIDI), funded by the Ministry of Health & Welfare, Korea (HI21C0977, HR22C1605).

**Conflict of interest**

Prof. Park has financial interests in Tomocube Inc., a company that commercializes holotomography instruments. All the other authors declare no conflict of interest.

*Annotated References*

*#2 Yue et al., 2022, The study shows that lipid droplet levels in muscle stem cells influence their ability to self-renew and regenerate muscle tissue.*

*#4. Pillai S et al., 2022, In cancer cells, activation of the acid-sensing receptor OGR1 triggers lipid droplet accumulation as an adaptive response to acidic stress, and that inhibiting OGR1 impairs both cell growth under acidic conditions and tumor formation.*

*#39 Park J et al., 2023, Artificial intelligence approaches for quantitative phase imaging techniques*

*#40. Jo Y et al., 2022, Virtual staining of subcellular organelles including lipid dropelts and mitochondria exploiting holotomography and 3D convolutional network*

*#43. Papsdorf K, et al., 2023, dietary mono-unsaturated fatty acids extend lifespan in Caenorhabditis elegans by upregulating both lipid droplets and peroxisomes, suggesting a co-regulated organelle network that modulates lipid homeostasis and longevity.*

*#46. Oh et al., 2022, The study introduces normalized Raman imaging, a microscopy technique that quantitatively measures protein and lipid concentrations in cells. This study also demonstrated the cross validation between normalized Raman imaging and holotomography.*

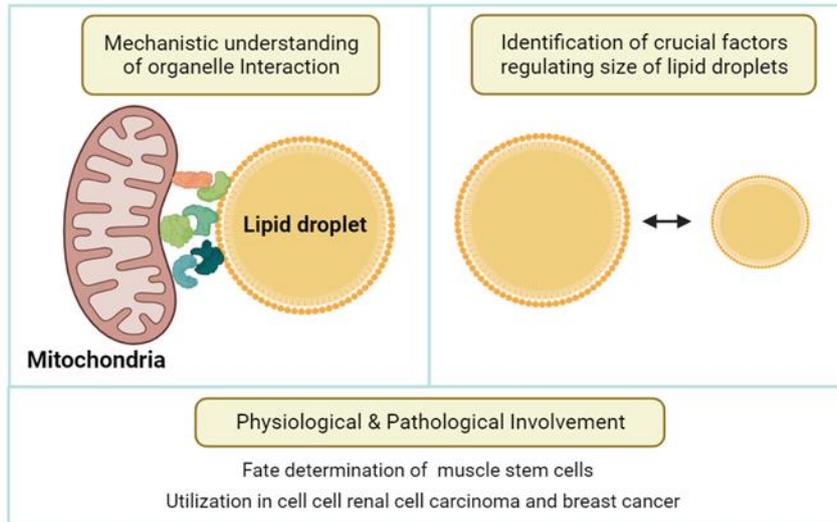

Figure 1: to overview the recent discovery of lipid droplets

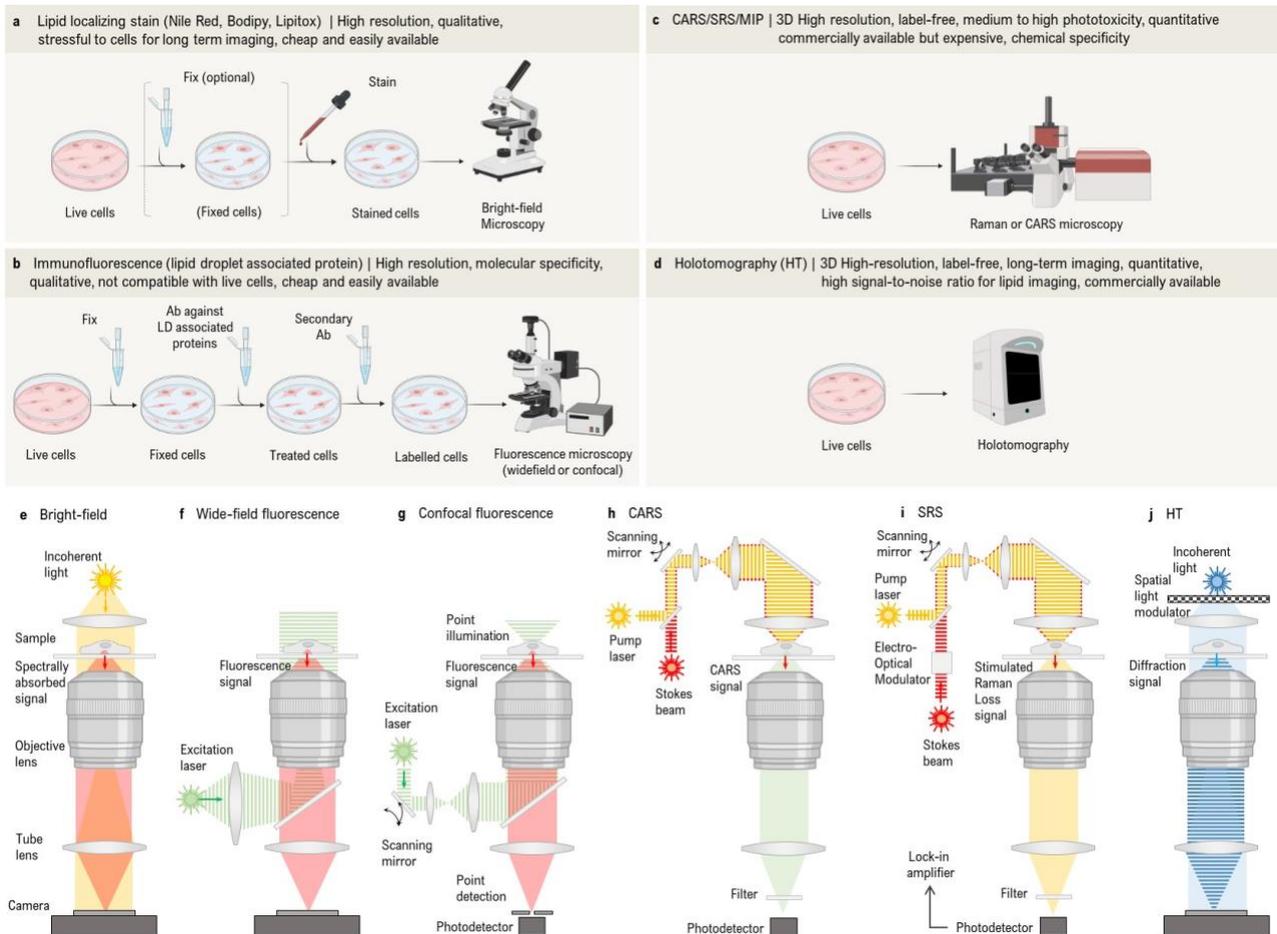

Figure 2. Work processes and schematics to show the principles lipid droplet imaging techniques. a | work process of lipid localizing staining methods. b | of immunofluorescence techniques. c | CARS/SRS/MIP techniques. d | holotomography technique. e-j | Schematics of imaging techniques for (e) bright-field microscopy, (f) wide-field fluorescence microscopy, (g) confocal fluorescence microscopy, (h) CARS microscopy, (i) SRS microscopy, and (h) HT.

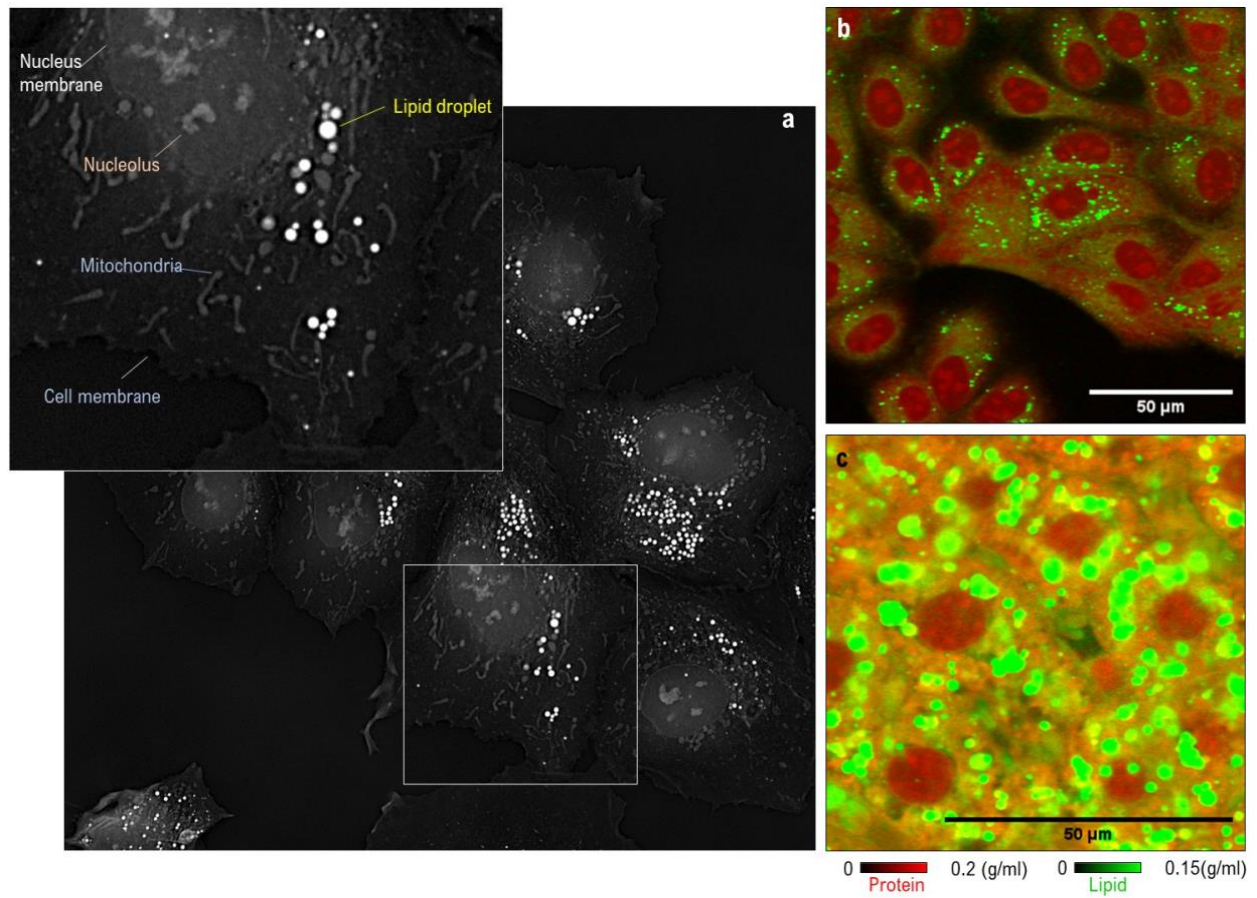

Figure 3. Representative label-free images of lipid droplets in live cells. a | HT images of lipid droplets in live unlabeled Hep3B cells. b-c | SRS images of (b) live unlabeled MDCK cells and (c) a fixed unlabeled mouse liver tissue.